\documentclass[preprint,authoryear,12pt]{elsarticle}

\usepackage{epsfig}

\usepackage{amssymb}

\usepackage[ps2pdf,%
a4paper=true,%
breaklinks=true,%
colorlinks=true,%
pdfauthor={Emil Khalisi et al. 2014},%
pdftitle={Counter Data of the Cosmic Dust Analyzer...}%
]{hyperref}

\journal{Advances in Space Research}

\usepackage{rotating}
\usepackage{lscape}
\usepackage[british]{babel}   

\begin{document}

\begin{frontmatter}



\title{Counter Data of the Cosmic Dust Analyzer aboard the
  {\it Cassini} spacecraft and possible ``dust clouds'' at Saturn}


\author{Emil Khalisi\corref{cor}\fnref{footnote2}}
\address{Max-Planck-Institute for Nuclear Physics, Saupfercheckweg 1, D--69117 Heidelberg}
\cortext[cor]{Corresponding author}
\ead{emil.khalisi@mpi-hd.mpg.de}


\author{Ralf Srama}
\address{Institute  for Space Systems, Pfaffenwaldring 29, D--70569 Stuttgart}

\author{Eberhard Gr\"un}
\address{Max-Planck-Institute for Nuclear Physics, Saupfercheckweg 1, D--69117 Heidelberg}

\begin{abstract}

We present the impact rates of dust particles recorded
by the Cosmic Dust Analyzer (CDA) aboard the {\it Cassini}
spacecraft.
The ``dust counters'' evaluate the quality of an impact
and give rise to the apparent density of dust particles
in space.

The raw data is pre-selected and refined to a new
structure that serves to a better investigation of
densities, flows, and properties of interplanetary
dust grains.
Our data is corrected for the dead time of the instrument
and corresponds to an assumed Kepler orbit (pointing of
the sensitive area).
The processed data are published on the website for the
Magnetosphere and Plasma Science (MAPSview), where it
can be correlated with other {\it Cassini} instruments.

A sample is presented for the Titan flyby on DOY 250/2006.
We find that the dust density peaks at two times,
at least, in a void region between Titan and Rhea.
Such features may point to extended clouds of small
particles drifting slowly in space.
These density clouds seem to be stable for as long as
several months or few years before dispersing.

\end{abstract}


\end{frontmatter}


\parindent=0.5 cm

\section{Introduction} \label{intro}

\noindent
The Cosmic Dust Analyzer (CDA) aboard the {\it Cassini}
spacecraft is an instrument to study the physical and
chemical properties of dust particles in the
interplanetary space \citep{srama-etal_2004}.
It consists of two subsystems working separately:
\begin{enumerate}
\item The main sensor is the Dust Analyzer (DA) designed to
analyse the particles according to their speed, mass,
charge, chemistry, and impact direction.
It deploys a time-of-flight mass spectrometer.
The sensitive area of the Impact Ionisation Detector (IID)
is 0.0825 m$^2$,
and Chemical Analyzer Target (CAT) is 0.0073 m$^2$.
\item The High-Rate Detector (HRD) only counts the number
of impacts with a frequency of up to 10,000 counts per
second.
It is a simple impact trigger with an area of 60 cm$^2$.
\end{enumerate}

The general purpose is to quantify the dust density in the
interplanetary space during the spacecraft cruise as well
as the dust environment at Saturn.
A permanent mapping of all directions is not possible
because of the fixed position of the CDA on a 3-axis
stabilised spacecraft.
However, a designated turntable of the CDA allows some
re-directioning of the boresight (aperture axis).

We introduce the ``dust counters'' that serve to
investigate local densities, particle flows, and their
properties measured along the {\it Cassini} trajectory
\citep{khalisi-srama_2012}.
The original data (raw data) of the CDA is stored in a
MySQL database on a computer server at the
Max-Planck-Institute for Nuclear Physics in Heidelberg,
Germany, and
Institute for Space Systems at the University of Stuttgart.

We wrote an IDL code that unites similar features in the
pristine data.
Additionally, the code returns useful graphics or correlates
various physical parameters about the distribution of dust.
Our code does not dispense the original data but serves 
as a helping tool if a more concise format is desired,
e.g.\ for comparison with other instruments on {\it Cassini}.
That pre-processed data can be accessed in the
MAPSview database (Magnetosphere and Plasma Science) at
{\tt http://mapsview.engin.umich.edu/}.

\section{Description of the Impact Counters} \label{ch:definecnt}

\begin{figure}
  \centering
  \includegraphics[width=\textwidth]{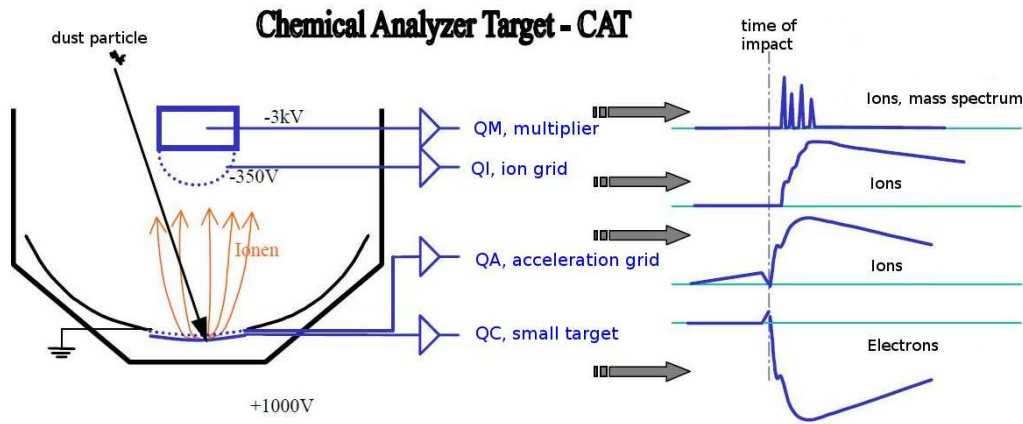}
  \caption{The QM signal provides a fast changing
      current that is based on the arrival times
      of chemical compounds of different mass
      \citep{srama_2000}.}
  \label{fig:qc-signal}
\end{figure}

\noindent
The ``dust counters'' are an array of 27 enumerators
for certain impact properties valid after DOY 150/2005.
Each event is analysed by the onboard flight software
immediately after its occurrence.
A dust grain hitting the sensitive area of the detector
will produce various signals that are subject to a
classification (Srama {\it et al.} 2006).
For example,
the QI-signal approximately indicates the impact energy
as returned by the ion multiplier,
while attributes like ``charge'' and ``velocity'' are exploited
by the QP-signal at the entrance grid (Fig.~\ref{fig:qc-signal}).
When spectral lines are present, the impact is supposed to
have hit the CAT rather than the IID.

The combined properties of all electronic signals are interpreted
by a decision making algorithm as such that the counters act
like ``containers'' or ``categories'' for some common
characteristics among the various dust particles.
The most important quantity for decision is the QI-signal as
given in column 3 of Table~\ref{tab:cnt-definitions} as
an example.

\begin{table}
\centering
\caption{Description of the counters (cnt) with their thresholds
for the ion signals (QI current) as well as digital numbers (dn);
priority sequence is indicated in the last column
(after \citet{srama_2000} and \citet{srama-etal_2006}).
The scheme is valid for impact events registered after DOY 150/2005.}
\label{tab:cnt-definitions}
\vspace{1.5ex}
\begin{tabular}{|c|c|rr|l|c|}
\hline
{\bf cnt} & {\bf name} &
         \multicolumn{2}{c|}{{\bf QI} [$\times 10^{-14}$ C]} &
         {\bf description} & {\bf prior.}  \\
\hline
 0 &  A0  & $> 320$  & (200 dn)  & CAT big, $>$3 mass lines   & 2 \\
 1 &  A1  & $> 166$  & (170 dn)  & CAT medium, $>$3 mass lines& 3 \\
 2 &  A2  & $> 67.1$ & (130 dn)  & CAT big,  2 mass lines     & 4 \\
 3 &  A3  & $> 33.1$ & (100 dn)  & CAT medium,  2 mass lines  & 5 \\
 4 &  A4  & $> 14.5$ & ( 70 dn)  & CAT small,  2 mass lines   & 15\\
 5 &  A5  & $> 5.82$ & ( 45 dn)  & CAT tiny,  2 mass lines    & 17\\
 6 &  A6  & $> 2.57$ & ( 30 dn)  & CAT small, no lines        & 18\\
 7 &  A7  & $> 1.03$ & ( 20 dn)  & CAT medium, relaxed        & 16\\
\hline
 8 &  I0  & $> 262$  & (190 dn)  & IID very big               & 8 \\
 9 &  I1  & $> 132$  & (160 dn)  & IID big + fast             & 9 \\
10 &  I2  & $> 67.1$ & (130 dn)  & IID medium + fast          & 10\\
11 &  I4  & $> 14.5$ & ( 70 dn)  & IID big + slow             & 11\\
12 &  I5  & $> 5.82$ & ( 45 dn)  & IID medium + slow          & 12\\
\hline
13 &  BO  &          &           & both targets (QC + QT)     & 22\\
14 &  N1  &\multicolumn{2}{l|}{Noise: QI flare}& signal on QI only & 1\\
15 &  S0  &          &           & spare: no function         & 23 \\
\hline
16 &  A8  & $> 0.0906$&( 12 dn)  & CAT tiny,  2 mass lines    & 20\\
\hline
17 &  I3  & $> 33.1$ & (100 dn)  & IID small + fast           & 13\\
18 &  I6  & $> 1.75$ & ( 25 dn)  & IID small + slow           & 14\\
19 &  I7  & $> 0.304$& ( 14 dn)  & IID tiny                   & 21\\
\hline
20 &  WO  & $> 1.75$ & ( 25 dn)  & Wall impact medium         & 6\\
21 &  W1  & $> 0.195$& ( 13 dn)  & Wall impact tiny           & 7\\
22 &  N0  &          &           & Noise: high baseline on QC & 19\\
23 &  N2  &          &           & Noise: any other signals   & 24\\
24 &  T0  &          &           & Test pulse: o.k.           & A1\\
25 &  T1  &          &           & Test pulse: wrong          & A2\\
26 &  E0  &          &           & internal interrupt counter & --\\
\hline
\end{tabular}
\end{table}

The algorithm is priority-sequenced (last column of
Tab.~\ref{tab:cnt-definitions}).
For example, the first check is whether the necessary 
signals are present at all, otherwise the event will be
classified as ``QI-flare'' (N1, see below).
Afterwards, the criteria for strong impacts on the CAT
are checked and whether a spectrum exists (A0 -- A3).
If not, then the event must have occurred elsewhere
(Wall, IID, etc.) but on the chemical target.
After having passed that onboard algorithm, each impact event
is assigned to one of the 27 counters (column 2)
and gets a qualitative name, e.g.\ ``big'' or ``slow''
(column 4).
A ``strong'' event can be caused by a massive but slow
particle as well as a small and very fast particle.
If the essential conditions of a special counter are met,
its value is enhanced by 1, and the search terminates.
If not, the next criteria are checked unless an appropriate
counter will be found.
Each impact is assigned to one unique counter.
The objective of that style is to accumulate the particles
with special features in order to compare their frequency
of occurrence.

The classification scheme follows a threefold pattern:
\begin{itemize}
\item A0 -- A8: counter names for the CAT.
      These are signals showing mass lines on the multiplier
      (QM signal) with a corresponding signal on the
      Chemical Analyser (QC signal);
      the impacts are further divided into 9 sub-classes
      (0 to 8) for large, medium, small responses as well as
      the number of line peaks on QM.
\item I0 -- I7: counter names for impacts on the IID;
      further division into 8 sub-classes (0 to 7)
      for fast and slow entrance velocities (QP signal
      at the entrance grid).
\item others: ten more counters are for putative noise
      records, impacts on the non-sensitive area (``wall''),
      test pulses, and control modes.
\end{itemize}

The thresholds for the registration of a particle were
changed throughout the mission to adjust for the
ambiences of a particular occasion:
dense regions, ring plane crossings, flybys at moons etc.
For example, when flying through a dense cloud, the number
of very small particles is high,
and large impacts are rarely triggered.
Then, the threshold was raised which resulted in a mode
of lower sensitivity of the CDA.
Therefore, the counters are a \emph{relative} measure at
different environments.
This kind of local customisation is the main bias that we
have to be aware of when using the data.

The counters were originally not meant for scientific use,
but rather to roughly prioritize the events for the
data readout from the spacecraft's repository.
The goal of altering thresholds was to select more important
dust records for a transfer to Earth than noise events
\citep{srama_2009}.
The concept of the counters turned out to be interesting
for correlations with properties that have been measured
by other {\it Cassini} instruments.

\section{Data Format for MAPSview} \label{dataformat}

\noindent
All counters are integer values.
They are read out on a semi-regular time period of
$dT =  t_j - t_{j-1} = 64$ seconds (standard interval).
Due to buffering effects and interrupts, this period
can occur as fractions (e.g.\ 29 + 35 s or 24 + 40 s).
However, multiples of 64 s can appear (128, 192, 256, etc.),
when the CDA encountered failure times,
or the downlink from the spacecraft was somehow affected.
We rectified all interrupts smaller than 64 sec to
ensure that no periods less than the ``standard'' is
present in our data.

\subsection{New counter groups}

\noindent
The original counter data of Table~\ref{tab:cnt-definitions}
are merged into 6 congeneric groups and later converted into
dust densities.
The six new groups are given in Table~\ref{tab:cntgroups}.
We distinguish between impacts on the ``CAT'' and ``IIT''.
The so-called ``wall'' events happen when only the QP-signal
at the entrance grid rises, but no others are noticeable;
such a particle seems to have entered the CDA hitting the
non-sensitive inner housing (wall).
``qi'' events are sudden flares on the ion collector
(QI) without accompanying signals elsewhere;
they cannot reliably be assigned to a particle impact.
Other signals of unknown origin, mostly weak ones, are
called ``noise''.
The last group, ``iitbig'', is a subgroup of ``iit'' and
receives special attention.

\begin{table}
\centering
\caption{Counters of Table~\ref{tab:cnt-definitions}
are merged to six congeneric groups.}
\label{tab:cntgroups}
\vspace{1.5ex}
\begin{tabular}{|l|l|}
\hline 
Group name  & 
                    Counters (after DOY 150/2005)\\
\hline \hline
noise       & 
                    B0 + N0 + N2    \\
wall        & 
                    W0 + W1         \\
cat         & 
                    A0 + A1 + A2 + A3 + A4 + A5 + A6 + A7 + A8 \\
iit         & 
                    I0 + I1 + I2 + I3 + I4 + I5 + I6 + I7 \\
iitbig      & 
                    I0 + I1 + I2 + I4 + I5 \\
qi          & 
                    N1              \\
\hline
all         & 
                    noise + wall + cat + iit + qi \\
\hline
\end{tabular}
\end{table}

We combined relevant counters in order to simplify the analysis.
Our goal is to investigate both a rougher mesh for the
multiple kinds of particles and a concise time scale with
accumulated impact events.
Micrometre-sized dust particles are still rare in the
environment of Saturn, therefore, it takes sometimes many
hours of no sole impact
(being regulated by the threshold for triggering).

\subsection{Impact rates}

\noindent
Since our counter groups contain integer values that may or
may not change within a standard time interval,
we converted them into impact rates $r$.
They are calculated by
\begin{equation}
   r = \frac{dN}{dT} = \frac{N_{j} - N_{j-1}}{t_j - t_{j-1}} , \label{eq:cntrate}
\end{equation}
where $N_j$ and $N_{j-1}$ are the particle numbers
in the corresponding interval of measurement $dT$.
The evaluation the impact event and storing takes up to
1 second which is called the ``dead time'' of the
instrument since it is insensitive to registrations of
other events.
This finite response time causes an error in the determination
of the true dust density in space:
In dense regions, when the instrument is expected to go
into saturation ($N_j - N_{j-1} \gtrsim 60$),
the impact rate will systematically be underestimated.

Since the statistics of dust impacts is based on the Poisson
process, a reconstruction of the ``true'' rate, $r^{\prime}$,
is possible as far as the statistical distribution of the
impact times is known.
The details of the correction procedure are discussed
by \citet{kempf_2008} and \citet{kempf_hb}.
To correct for the dead time of the impact rates,
we employ the simplified formula by \citet[his equation 3.43]{srama_2009}:
\begin{equation}
 r^{\prime} = \frac{r}{1 - \tau r} , \label{eq:ddt}
\end{equation}
where $\tau = 0.94$ s is a correction factor that emerges
from empirical tests as the most suitable compromise.
%
%
This formula yields almost identical results as the exact
solution by \citet{kempf_2008}.
Only if the instrument runs into saturation with impact
frequencies $\approx 0.98$ s$^{-1}$, the error exceeds 20\%.
The rates $r^{\prime}$ in equation~(\ref{eq:ddt}) are applied
to each counter group (noise, wall, cat, etc.) and
re-calculated at each time interval.
For example, to get the real rate, $r^{\prime}_{\rm iit}$,
of the dust impacts on the IIT at a given time,
we find $r^{\prime}_{\rm iit} = r_{\rm iit}/(1 - \tau r_{\rm iit})$.
A cross-check of our data samples for $r^{\prime}_{\rm all}$
showed an ample credibility with results obtained via
housekeeping data of the spacecraft.

\subsection{Sensitive area and dust ram}

\noindent
Two more parameters are necessary to obtain the spatial density
of the dust particles:
the sensitive area and the velocity of the spacecraft.
Since {\it Cassini} is stabilised in 3 axes,
the CDA has a field of view of $\pm 45^{\circ}$ depending on
the spacecraft's orientation in space (boresight).
However, the CDA has got an own pivoting table.
Its rotation is called ``articulation''
(0$^{\circ}$ to 270$^{\circ}$ with the zero-angle pointing
to the rear of the spacecraft).
It allows a re-alignment of the boresight to gain some
independence from the direction of flight.
The articulation is changed as needed, typically once or
twice per day.

\begin{figure}
  \centering
  \vspace{-0.5cm}
  \includegraphics[width=\textwidth]{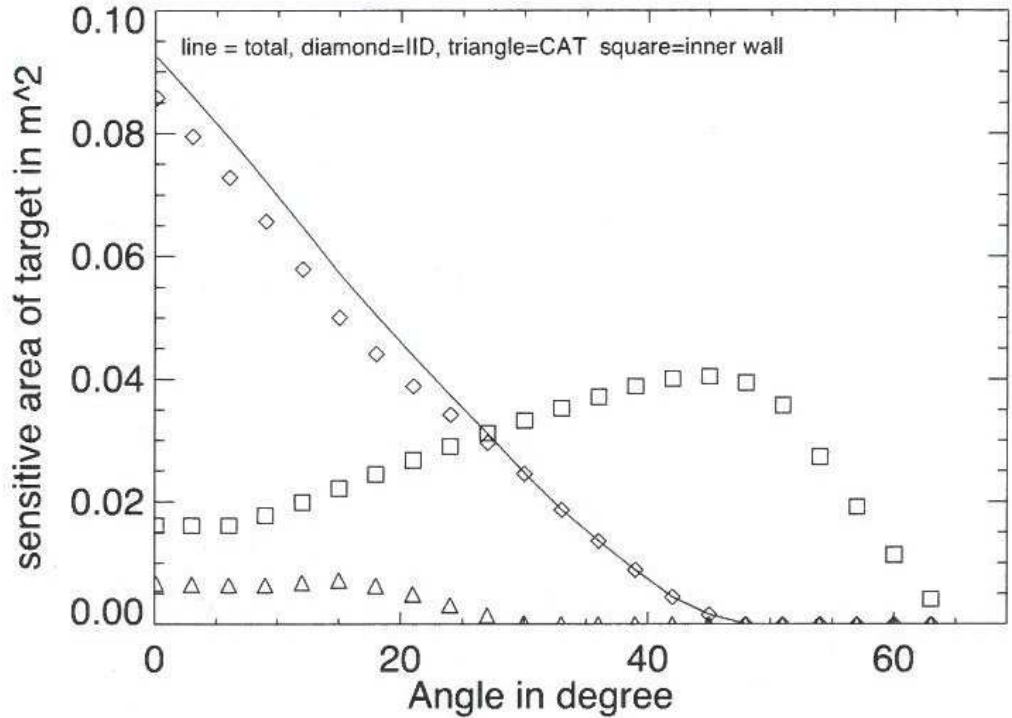}
  \caption{Sensitive area of the CDA as a function of the
     boresight:
     IID ($\diamond$), CAT ($\triangle$),
     and the non-sensitive housing ($\square$).
     The solid line indicates the sum of the measurements for
     IID and CAT \citep{srama_2009}.}
  \label{fig:sensarea}
\end{figure}

The sensitive area is shown in Figure~\ref{fig:sensarea}.
Calibration measurements with an isotropic flux of particles
yield a decrease of sensitivity for increasing incidence
angles.
Impacts onto the non-sensitive housing (``wall impacts'')
become predominant for angles $> 30^{\circ}$.
The measured rate is lowered when the axis of the instrument
is turned by the angle $\theta$, exibiting only its
``effective area'' to the particle flow.
It is simply:
$A_{\rm eff} = |\vec{A_0}| \cos(\theta)$,
with $\vec{A}_0$ being the total area of the CDA,
which is $|\vec{A}_0|$ = 0.09 m$^2$.
Obscuration by the ion collector in the focus was taken
into account.

\begin{figure}
  \centering
  \vspace{-0.7cm}
  \includegraphics[width=0.73\textwidth]{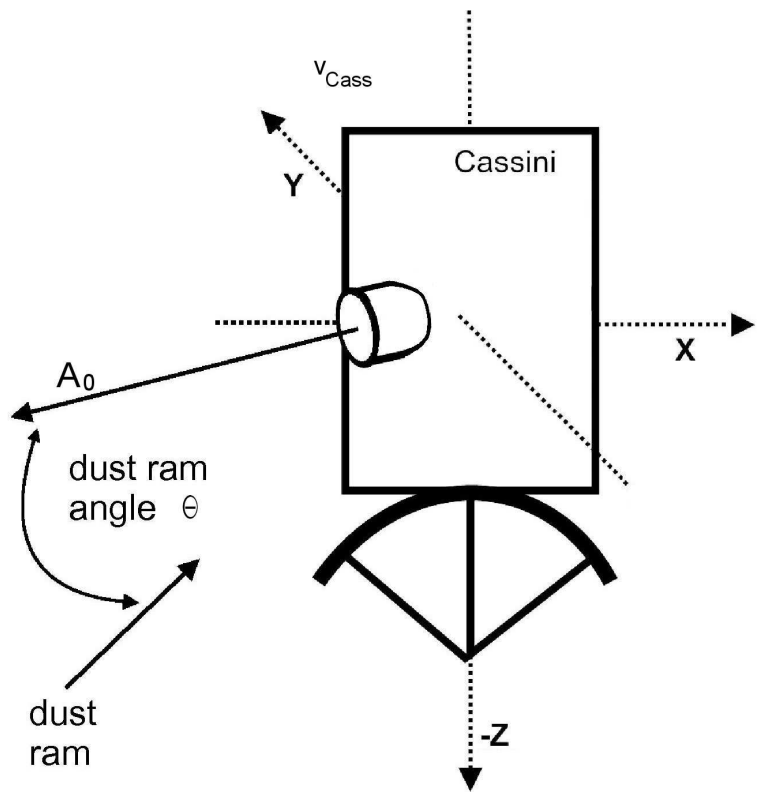}
  \caption{The pointing of the CDA may deviate from the
     dust ram and expose only a part of its
     sensitive area.}
  \label{fig:dustram3}
\end{figure}
Note, that the articulation angle of the CDA does \emph{not}
contain information about the direction of dust flow!
The particles can enter the CDA from any direction,
since they could be part of a stream with unknown origin,
or rubble piles, or stray bullets from a physical collision,
or interlopers from interstellar space etc.
Because of the lack of detailed information about the
planetologic conditions,
one assumption has to be made a priori, which is not
necessarily true:
The particle is supposed to have been on a circular
prograde orbit around Saturn before hitting the CDA.
That orbit marks the only trajectory to be reliably
calculated.
The angle between the boresight and the \emph{supposed}
flow of these particles is called the ``dust ram angle''
(Fig.~\ref{fig:dustram3}).
%
Thus, when the rate of wall impacts becomes abnormally
high or events are registered while $A_{\rm eff}$ is zero,
then the boresight of the CDA was aligned into a
non-Keplerian direction.

\subsection{Particle density}

\noindent
The velocity $\vec{v}_{\rm dust}$ of the particles can
be obtained from the difference of their Keplerian
velocity and the speed of the spacecraft $\vec{v}_{\rm Cass}$:
\begin{eqnarray}
   |\vec{v}_{\rm dust}| &=& |\vec{v}_{\rm Kepler} - \vec{v}_{\rm Cass}| \\
               &=& |\sqrt{GM_{\rm S}}\frac{\vec{a}}{|a|} - \vec{v}_{\rm Cass}| ,
\end{eqnarray}
where $G$ is the Gravitational constant,
$M_{\rm S}$ the mass of Saturn,
and $\vec{a}$ the distance of the particle from the
planet's center.
$\vec{v}_{\rm Cass}$ is given in the ephemeris data of
the {\it Cassini} mission.
If the modulus $|v_{\rm dust}|$ deviates much from the
particle velocity found through the QP-signal
(cf.\ Chapter \ref{ch:definecnt}), then the particle
would have been on a non-Keplerian orbit.

Finally, the local particle density $n$ is found as
\begin{equation}
n = \frac{dN}{dV}
  = \frac{r^{\prime}}{A_{\rm eff} \; |\vec{v}_{\rm dust}|},  \label{eq:density}
\end{equation}
where $dN$ denotes the number of particles in the
space volume $dV$;
$r^{\prime}$ is the dead time-corrected rate for
each counter group as in eq.~(\ref{eq:ddt})
within the interval $dT$.
In case of $A_{\rm eff} = 0$, the density $n$ was
set to zero.
Since the dust records for most of the counter groups
turn out to be rare events,
we chose to provide only $n_{\rm all}$ in our
Table~\ref{tab:datasample}.
Other sub-densities like $n_{\rm noise}$, $n_{\rm wall}$, etc.\
can be derived from eq.~(\ref{eq:density}) using the
corresponding $r^{\prime}_{\rm noise}$ and $r^{\prime}_{\rm wall}$,
respectively.

Table~\ref{tab:datasample} shows an excerpt of the resulting
matrix of impact rates and other parameters upon the alignment
of the CDA for 40 minutes on the day 251/2006.
We believe that such tables, structured for each day, comprehend
the vast amount of spacecraft data to the more relevant details.
The matrix is organised as follows:
\begin{itemize}
\item time of data readout, Julian Date, and the time leap $dT$,
      respectively;
\item $dN$: number of registered impacts in the time interval $dT$;
\item six columns for the particle rates of each impact
         group of Table~\ref{tab:cntgroups}:\\
      $r^{\prime}_{\rm noise}$: rate of noise events\\
      $r^{\prime}_{\rm wall}$: rate of wall impacts\\
      $r^{\prime}_{\rm cat}$: rate of impacts on the CAT\\
      $r^{\prime}_{\rm iit}$: rate of impacts on the IID\\
      $r^{\prime}_{\rm iitbig}$: rate of very prominent impacts on the IID
            (sub-quantity of $r^{\prime}_{\rm iit}$) \\
      $r^{\prime}_{\rm qi}$: rate of strong ``flares'' on the ion multiplier;
\item $r^{\prime}_{\rm all}$: the combined rate for all counters
        (without $r^{\prime}_{\rm iitbig}$);
\item $A_{\rm eff}$: sensitive area; $|\vec{v}_{\rm dust}|$: dust velocity;
      and $n$: particle density for all impacts registered ($r^{\prime}_{\rm all}$).
\end{itemize}
The complete set of our tables for the days 180/2005 to 109/2010
are publicly available at MAPSview (see Ch.~{\ref{intro}).

\begin{table}
\caption{Sample of the counter file
{\tt CDA$_{- -}$CNTS$_{-}$2006251.TAB}
that contains the particle rates and conditions of the CDA.
See text for details.
}  \label{tab:datasample}

\hspace{0.3cm}
\begin{sideways}
\begin{minipage}{0.95\textheight}
{\fontsize{6.2}{8} \selectfont
\begin{tabular}{p{1.6cm} p{1.4cm} p{0.15cm}p{0.25cm} *{7}{p{0.84cm}} p{0.8cm}p{0.8cm}p{0.6cm} }
Date/Time & Jul.Day & $dT$ & $dN$
   & $r^{\prime}_{\rm noise}$ & $r^{\prime}_{\rm wall}$ & $r^{\prime}_{\rm cat}$
   & $r^{\prime}_{\rm iit}$ & $r^{\prime}_{\rm iitbig}$ & $r^{\prime}_{\rm qi}$ & $r^{\prime}_{\rm al
l}$
   & $|\vec{A}_{\rm eff}|$ & $|\vec{v}_{\rm dust}|$ & $n$ \\[1ex]
\end{tabular}
\begin{verbatim}
2006-251T04:57:31  2453986.7066123   64   36   0.000000   0.000000   0.000000   0.000000   0.000000   1.193634   1.193634   0.07187760  6187.6689  2.6838E-03
2006-251T04:58:35  2453986.7073531   64   42   0.000000   0.000000   0.000000   0.000000   0.000000   1.712887   1.712887   0.07187760  6190.0418  3.8498E-03
2006-251T04:59:39  2453986.7080938   64   39   0.000000   0.000000   0.000000   0.036576   0.000000   1.389905   1.426481   0.07187760  6190.0418  3.2061E-03
2006-251T05:00:43  2453986.7088345   64   40   0.000000   0.000000   0.000000   0.000000   0.000000   1.515151   1.515151   0.07187760  6192.4173  3.4041E-03
2006-251T05:01:47  2453986.7095753   64   42   0.000000   0.000000   0.000000   0.040783   0.000000   1.672104   1.712887   0.07187760  6192.4173  3.8484E-03
2006-251T05:02:51  2453986.7103160   64   45   0.000000   0.000000   0.000000   0.000000   0.000000   2.073733   2.073733   0.07187760  6194.7936  4.6573E-03
2006-251T05:03:55  2453986.7110567   64   41   0.000000   0.039277   0.000000   0.000000   0.000000   1.571092   1.610369   0.07187760  6194.7936  3.6166E-03
2006-251T05:04:59  2453986.7117975   64   40   0.000000   0.037879   0.000000   0.000000   0.000000   1.477273   1.515151   0.06967440  6197.1740  3.5090E-03
2006-251T05:06:03  2453986.7125382   64   39   0.000000   0.000000   0.000000   0.000000   0.000000   1.426481   1.426481   0.06967440  6197.1740  3.3037E-03
2006-251T05:07:07  2453986.7132789   64   45   0.000000   0.046083   0.000000   0.000000   0.000000   2.027650   2.073733   0.06967440  6199.5570  4.8009E-03
2006-251T05:08:11  2453986.7140197   64   41   0.000000   0.000000   0.000000   0.039277   0.000000   1.571092   1.610369   0.06967440  6199.5570  3.7281E-03
2006-251T05:09:15  2453986.7147604   64   43   0.000000   0.042409   0.000000   0.000000   0.000000   1.781170   1.823579   0.06967440  6201.9426  4.2201E-03
2006-251T05:10:19  2453986.7155011   64   41   0.000000   0.000000   0.000000   0.000000   0.000000   1.610369   1.610369   0.06967440  6201.9426  3.7267E-03
2006-251T05:11:23  2453986.7162419   64   43   0.000000   0.000000   0.000000   0.000000   0.000000   1.823579   1.823579   0.06967440  6204.3307  4.2185E-03
2006-251T05:12:27  2453986.7169826   64   40   0.000000   0.000000   0.000000   0.037879   0.000000   1.477273   1.515151   0.06967440  6204.3307  3.5050E-03
2006-251T05:13:31  2453986.7177234   64   42   0.000000   0.000000   0.000000   0.000000   0.000000   1.712887   1.712887   0.06967440  6206.7213  3.9609E-03
2006-251T05:14:35  2453986.7184641   64   37   0.000000   0.000000   0.000000   0.000000   0.000000   1.266256   1.266256   0.06967440  6209.1144  2.9270E-03
2006-251T05:15:39  2453986.7192048   64   45   0.000000   0.000000   0.000000   0.000000   0.000000   2.073733   2.073733   0.06967440  6209.1144  4.7935E-03
2006-251T05:16:43  2453986.7199456   64   45   0.000000   0.000000   0.000000   0.000000   0.000000   2.073733   2.073733   0.06967440  6211.5101  4.7916E-03
2006-251T05:17:47  2453986.7206863   64   40   0.000000   0.000000   0.000000   0.000000   0.000000   1.515151   1.515151   0.06967440  6211.5101  3.5009E-03
2006-251T05:18:51  2453986.7214270   64   37   0.000000   0.000000   0.000000   0.000000   0.000000   1.266256   1.266256   0.07187760  6213.9083  2.8351E-03
2006-251T05:19:55  2453986.7221678   64   44   0.000000   0.000000   0.000000   0.000000   0.000000   1.943463   1.943463   0.07187760  6213.9083  4.3513E-03
2006-251T05:20:59  2453986.7229085   64   40   0.000000   0.000000   0.000000   0.000000   0.000000   1.515151   1.515151   0.07187760  6216.3074  3.3910E-03
2006-251T05:22:03  2453986.7236492   64   43   0.000000   0.000000   0.000000   0.000000   0.000000   1.823579   1.823579   0.07187760  6216.3074  4.0813E-03
2006-251T05:23:07  2453986.7243900   64   36   0.000000   0.000000   0.000000   0.000000   0.000000   1.193634   1.193634   0.07187760  6218.7107  2.6704E-03
2006-251T05:24:11  2453986.7251307   64   40   0.000000   0.037879   0.000000   0.000000   0.000000   1.477273   1.515151   0.07187760  6218.7107  3.3897E-03
2006-251T05:25:15  2453986.7258715   64   40   0.000000   0.000000   0.000000   0.000000   0.000000   1.515151   1.515151   0.07187760  6221.1167  3.3884E-03
2006-251T05:26:19  2453986.7266122   64   39   0.000000   0.000000   0.000000   0.000000   0.000000   1.426481   1.426481   0.07187760  6221.1167  3.1901E-03
2006-251T05:27:23  2453986.7273529   64   41   0.000000   0.039277   0.000000   0.000000   0.000000   1.571092   1.610369   0.07187760  6223.5251  3.5999E-03
2006-251T05:28:27  2453986.7280937   64   42   0.000000   0.000000   0.000000   0.000000   0.000000   1.712887   1.712887   0.07187760  6223.5251  3.8291E-03
2006-251T05:29:31  2453986.7288344   64   44   0.000000   0.000000   0.000000   0.000000   0.000000   1.943463   1.943463   0.07187760  6225.9361  4.3429E-03
2006-251T05:30:35  2453986.7295751   64   40   0.000000   0.000000   0.000000   0.000000   0.000000   1.515151   1.515151   0.07187760  6228.3496  3.3845E-03
2006-251T05:31:39  2453986.7303159   64   42   0.000000   0.000000   0.000000   0.000000   0.000000   1.712887   1.712887   0.07187760  6228.3496  3.8262E-03
2006-251T05:32:43  2453986.7310566   64   37   0.000000   0.000000   0.000000   0.000000   0.000000   1.266256   1.266256   0.07187760  6230.7657  2.8274E-03
2006-251T05:33:47  2453986.7317973   64   39   0.000000   0.000000   0.000000   0.000000   0.000000   1.426481   1.426481   0.07187760  6230.7657  3.1852E-03
2006-251T05:34:51  2453986.7325381   64   33   0.000000   0.000000   0.000000   0.000000   0.000000   1.000606   1.000606   0.07187760  6233.1843  2.2334E-03
2006-251T05:35:55  2453986.7332788   64   35   0.000000   0.032154   0.000000   0.000000   0.000000   1.093248   1.125402   0.07187760  6233.1843  2.5119E-03
2006-251T05:36:59  2453986.7340195   64   36   0.000000   0.000000   0.000000   0.000000   0.000000   1.193634   1.193634   0.07187760  6235.6055  2.6632E-03
2006-251T05:38:03  2453986.7347603   64   36   0.000000   0.000000   0.000000   0.000000   0.000000   1.193634   1.193634   0.07187760  6235.6055  2.6632E-03
2006-251T05:39:07  2453986.7355010   64   39   0.000000   0.000000   0.000000   0.000000   0.000000   1.426481   1.426481   0.07187760  6238.0276  3.1815E-03
2006-251T05:40:11  2453986.7362418   64   37   0.000000   0.000000   0.000000   0.000000   0.000000   1.266256   1.266256   0.07187760  6238.0276  2.8241E-03
2006-251T05:41:15  2453986.7369825   64   38   0.000000   0.000000   0.000000   0.000000   0.000000   1.343706   1.343706   0.07187760  6240.4538  2.9957E-03
\end{verbatim}
}
\end{minipage}
\end{sideways}
\end{table}

\section{Data Example at a Titan Flyby}

\noindent
A sample of the data is shown in Figure~\ref{fig:titanflyby}.
{\it Cassini} flew on its orbit no.\ 28 with an
inclination of 15$^{\circ}$ to the ring plane.
It came from the apokronium (DOY 240.80) aiming to the
inner regions of the Saturnian system.
From the middle panel it is visible that the aperture
axis of the CDA was not pointing to the dust ram till
DOY 250.83.
Nevertheless, the instrument registered a few particles
entering from any other direction, such that the
impact rate (upper panel) dithered at low values
$\approx 10^{-3}$ s$^{-1}$.
From 250.83 to 250.88 the justification of the sensitive
area changed according to the dust ram:
The whole probe was turned or the articulation was
modified, either of which will leave the same effect on the
impact rate:
it rose for a couple of minutes and vanished as sudden as
it appeared.

On DOY 250.84, {\it Cassini} experienced a targeted flyby
at Titan (T17) with a distance of 1000 km above its surface;
this flyby event perishes inside that first narrow peak.
Thereafter the sensitive area, $A_{\rm eff}$, became
better exhibited to the dust ram, and the dust rate rose
by 4 orders of magnitude.
%
\begin{figure}
  \caption{Impact rate, sensitive area, and the dust density
    during {\it Cassini}'s orbit no.~28 in the
    DOYs 250.3 -- 252.3/2006.
    A targeted Titan flyby happened at DOY 250.84.}
  \label{fig:titanflyby}
\vspace{0.2cm}
\hspace{0.8cm}
\begin{sideways}
  \includegraphics[width=1.35\textwidth]{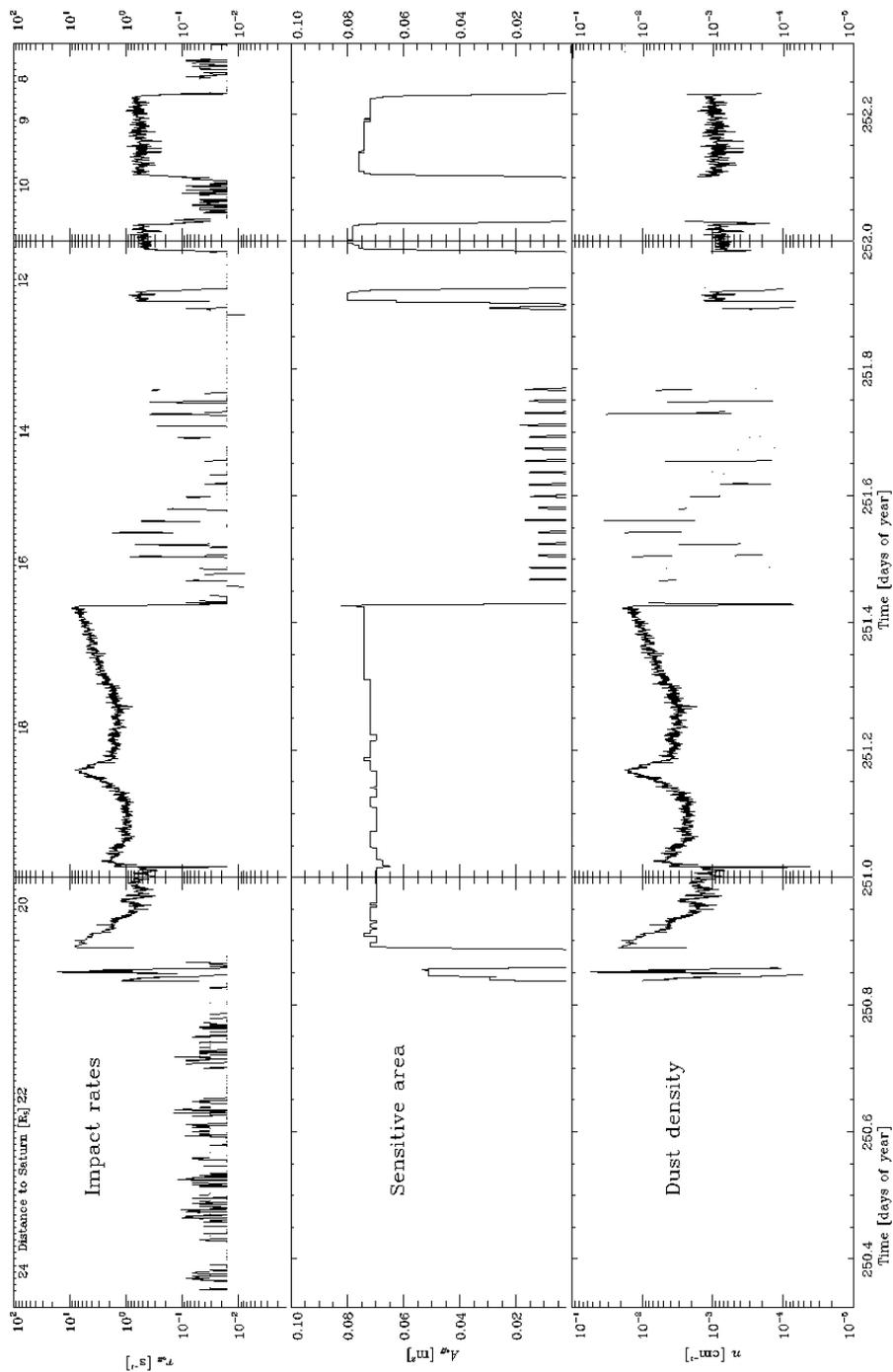}
%
%
\end{sideways}
\end{figure}

At DOY 250.89 the spacecraft crossed the ring plane from
North to South.
During such ring plane crossings the impact rate
naturally swells to a high level.
On its eccentric trajectory, {\it Cassini} headed for
Saturn:
It was to reach its perikronium at 3 R$_{\rm S}$ on
DOY 252.74 (not shown here).

Beyond DOY $\approx$ 251.42, the spacecraft began slowly
rotating along its $z$-axis, so,
the boresight of the CDA became alternately exposed and
hidden to the direction of the stream of particles.
That created the periodical peaks in the impact rate in
the upper panel (finishing at DOY $\approx$ 251.75).
The density (lower panel) reflects the interplay of
the oscillating impact rates and the lack of data
due to $A_{\rm eff} = 0$.

Two conspicuous enhancements of the particle density
are evident at DOYs 251.03 and 251.15, both of which
happened at distances between 19.7 and 18.7 R$_{\rm S}$,
respectively.
This is a quite empty region, for the orbit of inner
next moon Rhea ($d = 8.7$ R$_{\rm S}$) was still not
reached till the end of the day
(the moon itself was standing elsewhere on its orbit).
More detailed analysis showed that most of these hits
are ``N2 counts''
(see Table~\ref{tab:cnt-definitions}).
A third, very broad enhancement 
can also be presumed from the bottom panel.
It could have occurred at some time between DOY 251.50
and 251.70, after the beginning of the rotation of
the spacecraft.

\noindent
\begin{figure}
  \hspace{-0.5cm}
  \includegraphics[width=1.05\textwidth]{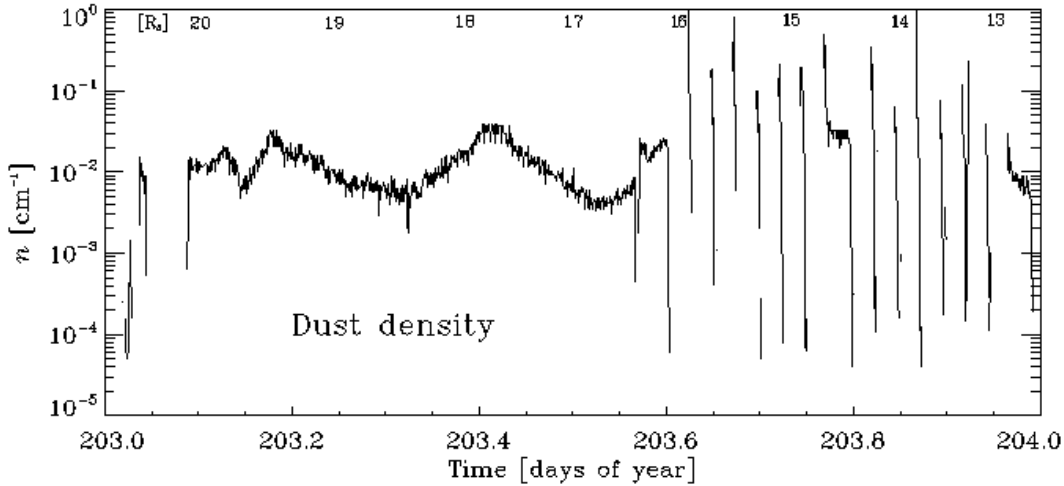}
  \caption{Particle density at {\it Cassini}'s orbit no.~26
    at DOY 203/2006.
    The numbers at the upper horizontal scale $[R_{\rm S}]$
    are the distance to Saturn.}
  \label{fig:orbit26}
\vspace{2ex}
%
\end{figure}

\section{Coherent dust clouds in the E-ring}

\noindent
The two or three enhancements of particles in
Figure~\ref{fig:titanflyby} could perhaps be attributed to
``dust clouds'' or any other compressions of dust that might
move in space.
In the region of their occurrence, the orbit of any known moon
is much too far for serving as an immediate source.
It is instructive to see that {\it Cassini} might have
flown through these ``clouds'' on its orbit no.\ 26.
From Figure~\ref{fig:orbit26} it is visible that the same
three enhancements (203.18, 203.40, and $\approx$ 203.8) 
were registered by the CDA.
They were positioned about 1 $R_{\rm S}$ closer to Saturn.
By simple velocity computation we derive an outward
drift of about 14 m/s.
The second enhancement peak appears ``widened'' which
might also be just an geometric effect of a slightly
different trajectory.

During the subsequent orbits no.\ 27 and no.\ 29
the boresight of the CDA was unfavorable positioned,
unfortunately, such that no measurements are available.
The velocity of the cloud particles is far too low for
stream particles that would originate from the planet
itself \citep{hsu_2011}.
However, our results will rather suit some less conspicuous
flows released at much earlier times and now slowly
floating through space.
The only known source of submicron material in the E-ring
is Enceladus at the radial distance of about 4 $R_{\rm S}$.

It might be speculative whether the clouds really originate
from that moon and move outwards now.
An implication of such an assumption will be that its
geysers are subject to different phases of activity.
Such temporary volcanos are only known from Jupiter's moon
Io, so far.
The possible particle clouds from Enceladus would
have to remain coherent for a long time, in this case
at hand as long as two years.

Alternative explanations are even more speculative in
our opinion, e.g.:
origins from dusty rings like Janus-, Pallene-,
G-ring, or even the F-ring that is heavily disturbed by
its shepherd moons.
Also, an origin from a dissolved comet is as unlikely as
raised ejecta from an impact on an inner moon.
To confirm the existence of such dusty clouds we look
forward to finding further evidence for similar
density enhancements at different locations.

\section*{Acknowledgments}

\noindent
E.K. wishes to thank the Klaus Tschira Foundation,
Heidelberg, for the kindful support at grant
no.~00.161.2010.
We also thank Sascha Kempf for using his Browser
software which enables to run this analysis,
as well as the referees for their invaluable comments
for improvement.


\end{document}